\documentclass[aps,pre,showpacs,superscriptaddress,10pt]{revtex4-1}
\pdfoutput=1
\usepackage[dvipsnames,rgb,dvips]{xcolor}
\usepackage{graphicx}
\usepackage{float}
\usepackage{overpic}
\usepackage{bbm}
\usepackage{amsmath}
\usepackage{amsfonts}
\makeatletter
\def\amsbb{\use@mathgroup \M@U \symAMSb}
\makeatother
\usepackage[bbgreekl]{mathbbol}
\usepackage{units,mathrsfs}

\newcommand\nn{\nonumber}
\newcommand\ve[1]{\boldsymbol{#1}}
\newcommand{\ma}[1]{\ensuremath{\amsbb{#1}}}

\newcommand{\Reys}{\ensuremath{\textrm{Re}_a}}

\newcommand{\St}{\ensuremath{\textrm{St}}}

\begin{document}
\title{Numerical analysis of the angular motion of a neutrally buoyant spheroid in shear flow at small Reynolds numbers}
\author{T. Ros\'en}
\affiliation{KTH Mechanics, Royal Institute of Technology, SE-100 44 Stockholm, Sweden}
\affiliation{Wallenberg Wood Science Center, Royal Institute of Technology, SE-100 44 Stockholm, Sweden}
\author{J. Einarsson}
\affiliation{Department of Physics, Gothenburg University, SE-41296 Gothenburg, Sweden}
\author{A. Nordmark}
\affiliation{KTH Mechanics, Royal Institute of Technology, SE-100 44 Stockholm, Sweden}
\author{C. K. Aidun}
\affiliation{George W. Woodruff School of Mechanical Engineering, and Parker H. Petit Institute for Bioengineering and Bioscience, 801 Ferst Drive, Georgia Institute of Technology, Atlanta, GA 30332-0405}
\author{F. Lundell}
\affiliation{KTH Mechanics, Royal Institute of Technology, SE-100 44 Stockholm, Sweden}
\affiliation{Wallenberg Wood Science Center, Royal Institute of Technology, SE-100 44 Stockholm, Sweden}
\author{B. Mehlig}
\affiliation{Department of Physics, Gothenburg University, SE-41296 Gothenburg, Sweden}

\begin{abstract}
We numerically analyse the rotation of a neutrally buoyant spheroid in
a shear flow at small shear Reynolds number. Using direct numerical stability analysis of the coupled nonlinear particle-flow problem we compute the linear stability of the log-rolling orbit at small shear Reynolds number, $\Reys$.
As $\Reys \to 0$ and as the box size of the system tends to infinity we find 
good agreement between the numerical results and earlier analytical predictions valid to linear order in $\Reys$ for the case of an unbounded shear. 
The numerical stability analysis indicates that there are substantial finite-size
corrections to the analytical results obtained for the unbounded system.
We also compare the analytical results to results of lattice-Boltzmann simulations to analyse the stability of the tumbling orbit at shear Reynolds numbers
of order unity. Theory for an unbounded system at infinitesimal shear Reynolds number predicts a bifurcation of the tumbling orbit at aspect ratio $\lambda_{\rm c} \approx 0.137$ below which tumbling is stable (as well as log rolling). 
The simulation results show a bifurcation line in the $\lambda$-$\Reys$ plane that reaches $\lambda \approx0.1275$ 
at the smallest shear Reynolds number ($\Reys=1$) at which we could simulate 
with the lattice-Boltzmann code, in qualitative agreement with the analytical results.
\end{abstract}

\maketitle

\section{Introduction}
The angular motion of a  neutrally buoyant spheroid in a simple shear has recently been studied extensively and in detail at moderately large shear Reynolds numbers, by numerical stability analysis and by computer simulations
using the lattice-Boltzmann method  \citep{Ding2000,qi2003,huang2012,rosen2014,mao2014,rosen2015a,rosen2015b}.  \citet{Ding2000} 
analysed rotation  in the flow-shear plane and
found that a saddle-node bifurcation gives rise to steady states where the symmetry axis of the particle aligns with a certain direction in the flow-shear plane. 
The authors of Refs.~\citenum{qi2003,huang2012,rosen2014,mao2014,rosen2015a,rosen2015b} analysed this bifurcation in detail  and found a large number of additional bifurcations at intermediate and large Reynolds numbers that give
rise to intricate angular dynamics.

At zero shear Reynolds number {by contrast} particle and fluid inertia are negligible, and the angular dynamics 
is determined by an infinite set of marginally stable periodic orbits, the so-called
Jeffery orbits \citep{Jeffery1922}. 

The effect of weak fluid and particle inertia on the angular motion of a neutrally buoyant spheroid  
in an unbounded shear
was analysed recently {using perturbation theory}~\cite{einarsson2014,einarsson2015a,einarsson2015b,candelier2015,subramanian2005}. In Refs.~\citenum{einarsson2015a,einarsson2015b} an approximate angular equation of motion was derived
for arbitrary aspect ratios of the spheroidal particle, {valid}
to linear order in the shear Reynolds number.  Linear stability analysis of
the Jeffery orbits subject to infinitesimal inertial perturbations
allowed to determine the linear stability of the log-rolling orbit (where
the particle symmetry axis is aligned with vorticity), and of tumbling
in the flow-shear plane: log rolling was found to be unstable for prolate spheroids and
stable for oblate spheroids, in agreement with the results obtained by \citet{subramanian2005} in the
slender-body limit. Refs.~\citenum{einarsson2015a,einarsson2015b} predicted that tumbling in the flow-shear plane is stable for prolate spheroids. For oblate spheroids tumbling was found to be stable for flat disks, otherwise  unstable. 

{This is a problem with a long history \cite{einarsson2015a}.} An earlier attempt \cite{saffman1956} to compute the stability of log rolling of nearly spherical particles {at infinitesimal shear Reynolds number} arrived at conclusions {at} variance with the results stated above, {namely that log rolling is stable for nearly spherical prolate spheroids. Lattice-Boltzmann simulations of the problem at moderate shear Reynolds number did not find stable log rolling for prolate spheroids \cite{qi2003,mao2014}, but as
pointed out in Ref.~\cite{mao2014} the shear Reynolds number was not small enough to allow for a definite comparison with theoretical predictions that consider the effect of fluid inertia as an infinitesimal perturbation.}

{This motivated us to} analyse the stability of the log-rolling orbit numerically at small shear Reynolds number {($\Reys$)} by discretising the coupled particle-flow problem directly. {This method is precise enough at sufficiently
small {$\Reys$} to determine which theory is correct, and for which values of the Reynolds number it applies.  
We find that the theory of Refs.~\citenum{einarsson2015a,einarsson2015b} 
agrees excellently with the simulation results at infinitesimal {$\Reys$} when the system size tends to infinity (the theory assumes that the shear is unbounded).}
{Our} numerical method allows to {compute the effect of confinement},
and to estimate the importance of higher-order $\Reys$-corrections to the analytical results {for the log-rolling orbit. To analyse} 
the bifurcations of the tumbling orbit at small shear Reynolds numbers
{we use lattice-Boltzmann} simulations.
At the smallest $\Reys$ attained with the lattice-Boltzmann code ($\Reys=1$)
the bifurcation occurs at a critical aspect 
ratio of $\lambda_{\rm c}\approx 0.1275$
in the finite system, in qualitative agreement with the analytical results obtained for an unbounded system.

We briefly comment on the wider context of this paper.
Recently there has been a surge of interest in describing the tumbling of small non-spherical particles in turbulent \cite{Marchioli2010,Par12,Pum11,Ni14,Che13,Challabotla2015,Byron2015} and complex flows \cite{Wil09,Wil10a,Wil11} using Jeffery's equation. Studies of the dynamics of larger non-spherical particles in turbulence \cite{Marchioli2010,Gus14,Challabotla2015} take into account particle inertia but neglect fluid inertia because it is difficult to solve the coupled particle-flow problem. For heavy particles this may be a good approximation, but the results summarised in this paper (and the results of Refs.~\citenum{subramanian2005,rosen2014,rosen2015a,rosen2015b,einarsson2015a,einarsson2015b,candelier2015}) show that this is approximation is likely to fail for neutrally buoyant and
nearly neutrally buoyant particles.

The remainder of this paper is organised as follows. Section \ref{sec:problem} describes
the coupled particle-flow problem that is the subject of this paper. 
In Section \ref{sec:theory} we summarise
the analytical results of Refs.~\citenum{einarsson2015a,einarsson2015b,candelier2015} and
find the bifurcations of the angular equation of motion obtained in these references.
Our numerical results are described in Section \ref{sec:numerics}, and compared
to the analytical results. Section \ref{sec:conc} contains our conclusions.

\section{Formulation of the problem}
\label{sec:problem}
The problem has the following dimensionless parameters. The shape of the spheroid is
determined by the shape
factor $\Lambda$ defined as $\Lambda = (\lambda^2-1)/(\lambda^2+1)$ where
$\lambda$ is aspect ratio of spheroid, $\lambda = a/b$ for prolate spheroids, $a$ is the major semi-axis length of the particle, and $b$ is the minor semi-axis length. For oblate spheroids the aspect ratio is defined as $\lambda = b/a$. The effect of fluid inertia is measured by the shear 
Reynolds number $\Reys = a^2s/\nu$ where $\nu$ is the kinematic
viscosity of the fluid and $s$ is the shear rate.
Particle inertia is measured by the Stokes number,
$\St= (\rho_{\rm p}/\rho_{\rm f})\Reys$ where $\rho_{\rm p}$ and $\rho_{\rm f}$
are particle and fluid mass densities. The numerical computations described in this paper
are performed in a finite system of linear size $L$, and $\kappa = 2a/L$ 
is a dimensionless measure of the system size,
 $2a$ is the length of the major axis of the particle.

We use dimensionless variables to formulate the problem.
{The length} scale is taken to be the major semi-axis length $a$ of the spheroid. The velocity
scale is $as$, the pressure scale is $\mu s$, and force and torque scales
are $\mu sa^2$ and $\mu s a^3$ respectively, $\mu$ is the dynamic viscosity
of the fluid.
In dimensionless variables the angular equations  of motion read
\begin{align}
\label{eq:particle_eom}
\dot {\ve n} &= \ve  \omega \wedge \ve n\,,\quad
\St\, \dot{\ve L} = \St\, (\ma I \dot{\ve \omega} + \dot{\ma  I} \ve \omega) = \ve T\,.
\end{align}
Here $\ve n$ is the unit vector along the particle symmetry axis, 
dots denote time derivatives, $\ma I = A^I (\mathbb{1} -\ma P_\perp) + B^I \ma P_\perp$
is the particle-inertia matrix,
$\ma P_\perp$ is a projector onto the plane perpendicular to $\ve n$
with elements $P_{ij} = \delta_{ij} - n_i n_j$, and $A^I$ and $B^I$ are moments of inertia
along and orthogonal to $\ve n$.
The particle angular velocity is $\ve \omega$, and $\ve T$ is the
hydrodynamic torque:
\begin{equation}
\label{eq:torque}
\ve T = \int_\mathscr{S}\!\! \ve r \wedge  \bbsigma {\rm d}\ve s \,.
\end{equation}
The integral is over the particle surface $\mathscr{S}${,} $\ve r$ is the position vector, and 
$\bbsigma$ is the stress tensor with elements $\sigma_{ij} =   -p\delta_{ij} +  
2S_{ij}$ where $p$ is pressure, and $S_{ij}$ are the elements of the strain-rate matrix $\ma S$, 
the symmetric part of the matrix $\ma A$ of fluid-velocity gradients
with elements $A_{ij} = \partial_j u_i$ 
($u_i$ are the components of the fluid velocity $\ve u$).
The anti-symmetric part of $\ma A$ is denoted by  $\ma O$
with elements $O_{ij}$.
To determine the torque it is necessary to
solve the Navier-Stokes equations for the incompressible fluid:
\begin{equation}
\label{eq:ns}
\Reys \big( \partial_t \ve u + (\ve u\cdot \ve\nabla)\ve u\big) = -\ve \nabla  p + 
\Delta \ve u\,,\quad  \nabla \cdot \ve u = 0\,.
\end{equation}
For a neutrally buoyant particle $\Reys=\St$. 

\begin{figure}[t]
\includegraphics[width=10cm,clip]{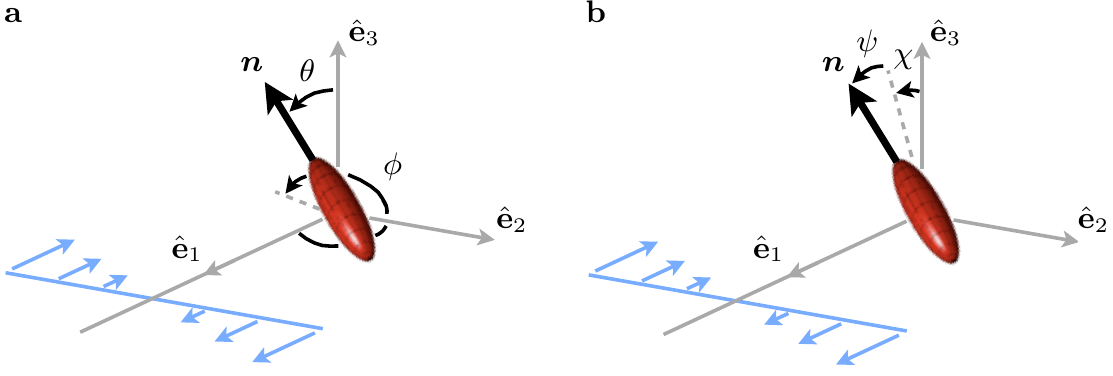}
\caption{\label{fig:1} {(Colour online)}.
Schematic illustration of spheroid in a simple shear in a 
coordinate system that translates with centre of mass of the particle.
Vorticity points
along the negative $\hat{\bf e}_3$-axis, and $\hat{\bf e}_1$ is the flow direction. The flow-shear plane is spanned by  $\hat{\bf e}_1$ and  $\hat{\bf e}_2$. We use two different coordinate
systems to express the orientation of the unit vector $\ve n$ aligned with the symmetry axis of the particle. {\bf a} Spherical coordinate
system used for analysing linear stability of tumbling in the flow-shear plane,
$\theta$ is the polar angle from the vorticity axis,
and $\phi$ is the azimuthal angle in the flow-shear plane.
{\bf b} Spherical coordinate system used for analysing linear stability of log rolling,
$\ve n = [0,0,1]$ corresponds to $\chi=\psi=0$.
}
\end{figure}

It is assumed that the slip velocity vanishes
on the particle surface $\mathscr{S}$, 
$\ve u\!=\!\ve \omega \wedge \ve r \quad\mbox{when}\quad\ve r\! \in \! \mathscr{S}$. The perturbation calculations in
Refs.~\citenum{einarsson2015a,einarsson2015b,candelier2015} apply to a simple
shear in an unbounded system, and it is assumed that the fluid velocity 
far from the particle is unaffected by its presence: 
$\ve u\! =\! \ve u^\infty$ as $|\ve r|\!\to\! \infty$. Here $\ve u^\infty$ 
denotes the velocity field of the simple shear
$\ve u^\infty = \ma A^\infty\ve r$  with $A^\infty_{ij} = \delta_{i1}\delta_{j2}$ (see Fig.~\ref{fig:1} for an illustration of the geometry).
The symmetric and antisymmetric parts of $\ma A^\infty$ are denoted
by $\ma S^\infty$ and $\ma O^\infty$, respectively.

The numerical computations described in this paper pertain to a finite system, a cube of linear size $2\kappa^{-1}$ (in dimensionless variables).  In the shear direction $u_1 = \pm \kappa^{-1}$ at $r_2 = \pm \kappa^{-1}$. 
In the flow  and vorticity directions periodic boundary conditions are used.

\section{Theory at small $\Reys$.}
\label{sec:theory}

In Refs.~\citenum{einarsson2015a,einarsson2015b,candelier2015} an approximate angular
equation of motion for a neutrally buoyant spheroid in an unbounded shear flow was derived,  valid to linear order in $\Reys=\St$:
\begin{align}
\label{eq:eom_symmetry}
  \dot {\ve n} \!&=\!
\ma O^\infty \ve n
\!+\! \Lambda[\ma S^\infty\ve n\! -\! (\ve n \cdot \ma S^\infty \ve n)\ve n] 
+\beta_1 (\ve n \cdot \ma S^\infty \ve n)\ma P_\perp\, \ma S^\infty \ve n\!\\
&+\! \beta_2 (\ve n \cdot \ma S^\infty \ve n)\ma O^\infty \ve n +\beta_3\, \ma P_\perp\, \ma O^\infty\, \ma S^\infty \ve n \!
+\! \beta_4\, \ma P_\perp\, \ma S^\infty\, \ma S^\infty \ve n\,. \nn
\end{align}
The first two terms on the r.h.s. of this equation are Jeffery's result for a neutrally buoyant spheroid in the creeping-flow limit. The remaining terms are corrections due to particle and fluid inertia.  The four coefficients $\beta_\alpha$ (for $\alpha = 1,\ldots,4$) are
linear in $\Reys$ and $\St$ but non-linear
functions of the particle aspect ratio $\lambda$: $\beta_\alpha=b_\alpha^{(\Reys)}(\lambda)\Reys+b_\alpha^{(\St)}(\lambda)\St$.
These functions were computed by \citet{einarsson2015a,einarsson2015b}
for general values of $\lambda$, and in Ref.~\citenum{candelier2015} in the nearly-spherical
limit. Eq. (\ref{eq:eom_symmetry}) determines the effect of small inertial perturbations on
the Jeffery orbits. It turns out that log rolling ($\ve n$ aligned with the vorticity axis)
and tumbling in the flow-shear plane survive small inertial perturbations. In the following
two Sections we discuss the linear stabilities of these two orbits, for $\St=\Reys$.
We write $\beta_\alpha = \Reys\, b_\alpha(\lambda)$. 
Table~\ref{tab:1} gives the asymptotes of these functions  
for large and small values of the aspect ratio $\lambda$. The asymptotes are obtained
by expanding the results derived in Ref.~\citenum{einarsson2015b}.

\subsection{Linear stability analysis of log rolling}
To analyse the stability of the log-rolling orbit we use the coordinate
system shown in Fig.~\ref{fig:1}{\bf b}. The angles 
$\chi$ and $\psi$ are defined so that
\begin{equation}
n_1 = \sin\psi\,,\quad n_2 = \cos\psi \sin\chi\,,\quad n_3 = \cos\psi\cos\chi\,.
\end{equation}
In these coordinates  the equation of motion (\ref{eq:eom_symmetry}) takes the form:
\begin{subequations}
\label{eqn:psi}
\begin{align}
  \dot \psi  & =\frac{1}{8} \big[4(\Lambda \cos 2\psi+1)\sec\psi\sin\chi \\
             & \hspace*{3mm}+\big(4\beta_1\cos2\psi\sin^2\chi+(-2\beta_2-\beta_3+\beta_4)\cos2\chi
               +2\beta_2 + 3\beta_3 +\beta_4\big)\sin\psi\big]\cos\psi\,,\nn\\
 \dot \chi &= \frac{1}{4}\big[2(\Lambda-1)\tan\psi+\big((\beta_2-\beta_1)\cos2\psi + \beta_1-\beta_2-\beta_3+\beta_4\big)\sin\chi\big]\cos\chi\,.
\end{align}
\end{subequations}
Log rolling along the vorticity
direction $\ve n = [0,0,1]$ corresponds to $\chi=\psi=0$, 
and this is a fixed point of Eq. (\ref{eqn:psi}) since $\dot \psi = \dot \chi = 0$
in this direction. The stability of this fixed point is determined by 
the eigenvalues of the linearisation of Eq.~(\ref{eqn:psi}) 
around this fixed point. To linear order in $\Reys$ the eigenvalues take the form
\begin{equation}
\label{eq:gammaLR}
\gamma_{\rm LR}^\pm = \frac{\beta_4}{4} \pm \frac{\rm i}{2} \sqrt{1-\Lambda^2}+o(\Reys)\,.
\end{equation}
The real part of this expression was derived in Refs.~\citenum{einarsson2015a,einarsson2015b} (see for example Fig.~3{\bf a} in Ref.~\citenum{einarsson2015a}). The coefficient $\beta_4$  is linear in $\Reys$ and its sign
determines the stability of the log-rolling orbit at infinitesimal $\Reys$.
The coefficient is positive for prolate spheroids
(unstable log rolling) and negative for oblate spheroids (stable log rolling).
The imaginary part in Eq.~(\ref{eq:gammaLR}) shows that the log-rolling
fixed point is a spiral at small $\Reys$. 
The imaginary part has no correction to linear order in $\Reys$.

\begin{table}
\caption{\label{tab:1} Asymptotic behaviour of the functions
$b_\alpha(\lambda )= \beta_\alpha/\Reys$ where $\beta_\alpha$ are the coefficients in
Eq. (\ref{eq:eom_symmetry}) for $\St=\Reys$. The asymptotes are found by
expanding the  solutions from Refs.~\citenum{einarsson2015a,einarsson2015b}. }
\mbox{}\\
\begin{tabular}{lcc}
\hline\hline
 & \hspace*{2cm}\mbox{prolate}\hspace*{2cm} &\hspace*{2cm} \mbox{oblate}\hspace*{2cm}\\[1mm]
 & \hspace*{2cm}$\lambda \to \infty$\hspace*{2cm} & \hspace*{2cm}$\lambda\to 0$\hspace*{2cm}\\[1mm]\hline\\[-3mm]
$b_1$ &$\frac{7}{15(2\log\lambda-3+\log4)}
 \!+\!\frac{-197\log2\lambda+92\log\lambda\log4\lambda+106+92(\log2)^2}{15\lambda^2(2\log\lambda-3+\log4)^2}$
      & $\frac{11}{30}\!+\!\big(\frac{176}{45\pi}\!-\!\frac{7\pi}{20}\big)\lambda\!+\!\big(-\frac{7}{3}\!+\!\frac{3968}{135\pi^2}\!-\!\frac{21\pi^2}{80}\big)\lambda^2$\\[2mm]
$b_2$ &
 $\frac{1}{5(2\log\lambda-3+\log4)}\!+\!\frac{(\log\lambda-1+\log2)(8\log2\lambda-7)}{5\lambda^2(2\log\lambda-3+\log4)^2}$
      &$\frac{1}{10}\!+\!\big(\frac{8}{15\pi}\!-\!\frac{\pi}{20}\big)\lambda
        \!+\!\big(-\frac{1}{5}\!+\!\frac{128}{45\pi^2}\!-\!\frac{3\pi^2}{80}\big)\lambda^2$
\\[2mm]
$b_3$ &$-\frac{4}{5\lambda^2}$
      &$-\frac{1}{5} \!+\!\frac{9\pi^2-64}{60\pi}\lambda
      \!+\!\big(\frac{3}{5}-\frac{256}{45\pi^2}\!+\!\frac{9\pi^2}{80}\big)\lambda^2$\\[2mm]
$b_4$ &$\frac{4}{15\lambda^2}$&$ -\frac{1}{3}\! +\!\big(\frac{\pi }{20}\!-\!\frac{64}{45 \pi }\big) \lambda\! +\!\big(\frac{5}{3}\!-\!\frac{1024}{135 \pi ^2}\!+\!\frac{3 \pi ^2}{80}\big) \lambda ^2$\\[2mm]
\hline\hline
\end{tabular}
\end{table}

\subsection{Tumbling in the flow-shear plane}
Under which circumstances is tumbling in the shear plane stable? 
In this Section we first
summarise the results of analytical linear-stability calculations of
Refs.~\citenum{einarsson2015a,einarsson2015b,candelier2015}
at infinitesimal $\Reys$. Second we discuss finite but small shear Reynolds numbers.
To analyse tumbling in the flow-shear plane we use the
coordinates employed in Refs.~\citenum{einarsson2015a,einarsson2015b,candelier2015}
(illustrated in Fig.~\ref{fig:1}{\bf a}):
\begin{equation}
n_1 = \sin\theta\cos\phi\,,\quad n_2 = \sin\theta\sin\phi\,,\quad n_3 = \cos\theta\,.
\end{equation}
In these coordinates the equation of motion (\ref{eq:eom_symmetry}) takes the form
\begin{subequations}\label{eq:effective_spherical}
\begin{align}
  \dot \phi &=\frac{1}{2}\left(\Lambda  \cos 2 \phi -1\right) +  \frac{1}{8} \beta_1 \sin ^2\theta \sin 4 \phi -\frac{1}{4} \sin 2 \phi \left(\beta_2 \sin ^2\theta+\beta_3\right)\,,
  \\
  \dot \theta &= \Lambda  \sin \theta  \cos \theta \sin \phi  \cos \phi + \frac{1}{4} \sin \theta \cos \theta \left(\beta_1 \sin ^2\theta \sin ^2 2\phi+\beta_3 \cos 2 \phi +\beta_4\right)\,. 
\label{eq:dottheta}
\end{align}
\end{subequations}
This is Eq.~(42) in Ref.~\citenum{einarsson2015a}. Eq. (\ref{eq:dottheta}) shows
that $\dot\theta=0$ at $\theta=\pi/2$, in the flow-shear plane. The equation
of motion for $\phi$ in this plane is 
\begin{equation}
\dot\phi = \frac{1}{2}\left(\Lambda  \cos 2 \phi -1\right)+ \frac{1}{8} \beta_1\sin4\phi
-\frac{1}{4} (\beta_2+\beta_3)\sin2\phi\,.
\end{equation}
At infinitesimal values of $\Reys$ there is a periodic tumbling orbit
in the flow-shear plane because $\dot\phi < 0$. Its linear stability exponent $\gamma_{\rm T}$
at infinitesimal shear Reynolds numbers was calculated in Refs.~\citenum{einarsson2015a,einarsson2015b}. It was found that tumbling in the flow-shear plane is stable for prolate particles in this limit, and unstable for not too thin oblate particles. 
For thin platelets tumbling was found to be stable. For infinitesimal
shear Reynolds numbers the bifurcation occurs at the critical aspect
ratio  \cite{einarsson2015a,einarsson2015b} 
 \begin{equation}
 \label{eq:lambdac}
 \lambda_{\rm c} \approx 0.137\,.
 \end{equation}
This concludes our summary of the results of Refs.~\citenum{einarsson2015a,einarsson2015b}, valid at infinitesimal $\Reys$. 

As $\Reys$ increases we see that $\dot\phi\geq 0$ in Eq.~(\ref{eq:effective_spherical}) for some value(s) of $\phi$. This implies the existence of fixed points in the flow-shear plane. 
This happens in Eq.~(\ref{eq:effective_spherical}) for any aspect ratio. 
But Eq.~(\ref{eq:effective_spherical}) is valid only to linear order in $\Reys$. For this reason we only look
at limiting cases where Eq.~(\ref{eq:effective_spherical}) exhibits bifurcations at small values of  $\Reys$.
This occurs for thin rods and plates, as will be seen below.

Consider first  rods. Rods of infinite
aspect ratio align with the flow direction, particles with
finite aspect ratio tumble at infinitesimal $\Reys$.
At finite values of $\Reys$ a bifurcation may cause a rod  with finite aspect ratio to align. To find      
this bifurcation point we expand $\dot\phi$ to second order in 
$1/\lambda$ (Table~\ref{tab:1}) and to second order in $\phi$. 
A double root of the resulting quadratic equation for $\phi$ determines the 
bifurcation point:
\begin{equation}
\label{eq:Rec2}
\Reys^{({\rm c1})} \sim
\frac{15}{\lambda}\big(-3+\log 4 + 2\log \lambda\big)\quad\mbox{as $\lambda\to\infty$}\,.
\end{equation}
The leading terms of this result for $\Reys^{({\rm c1})}$
agree with Eq.~(3.31) in Ref.~\citenum{subramanian2005} (up to a factor of $8\pi$).
\citet{subramanian2005} derived their result
using the slender-body approximation.
Note  that the qualitative features of the dynamics
in the vicinity of $\Reys^{({\rm c1})}$ are consistent with Eq.~(12) in Ref.~\citenum{Ding2000}
(see also \citet{Zettner2001}). As $\varepsilon \equiv \Reys- \Reys^{({\rm c1})} $ tends to zero 
from below the period of the tumbling orbit tends to infinity as $(-\varepsilon)^{-1/2}$. Above the transition there are two fixed points, a saddle point and a stable node. It follows that the particle aligns at the angle 
\begin{equation}
\label{eq:phi_prolate}
\phi_0 =  \frac{1}{{\lambda}}+ 
 \frac{\sqrt{\varepsilon}}{15}  \frac{\sqrt{30}}{\sqrt{\lambda(-3+\log 4+2\log\lambda)}}+\ldots
\quad\mbox{as $\lambda\to\infty$,}
\end{equation}
for small values of $\varepsilon$. The form of this equation is 
consistent with Eqs.~(3.30) and ~(3.31) in Ref.~\citenum{subramanian2005}.

Now we turn to thin disks.  The symmetry vector
of an infinitely thin disk aligns with the shear direction, $\dot\phi=0$ for $\phi=\pi/2$ when $\lambda=0$. For non-zero values of $\lambda$ the vector $\ve n$  tumbles in the flow-shear plane in the limit of $\Reys\to 0$. At finite (but small)
values of $\Reys$ a bifurcation may cause the disk to align. To find
this bifurcation point we expand $\dot\phi$ to second order in
$\lambda$ (Table~\ref{tab:1}) and to second order in $\delta\phi=\phi-\pi/2$.
As above a double root of the resulting quadratic equation for $\delta\phi$ determines 
the critical shear Reynolds number:
\begin{equation}
\label{eq:Rec3}
\Reys^{({\rm c2})} \sim 
 15 \lambda 
\quad\mbox{as $\lambda\to0$}\,.
\end{equation}
For $\Reys>\Reys^{({\rm c2})}$ the symmetry vector $\ve n$ of the disk
aligns in the flow-shear plane at the angle
\begin{equation}
\label{eq:phi_oblate}
\phi_0 = \frac{\pi}{2} +\lambda + \frac{\sqrt{\epsilon}}{15}\, \sqrt{30 \lambda} \quad\mbox{as $\lambda\to 0$.}
\end{equation}
In deriving this expression only the lowest orders 
in $\lambda$ and $\epsilon$ were kept.

The bifurcation lines in the $\lambda$-$\Reys$-plane given
by Eqs.~(\ref{eq:lambdac}), (\ref{eq:Rec2}), 
and (\ref{eq:Rec3}) are shown in Fig.~\ref{fig:lambdac}. This figure also contains
the results of our direct numerical simulations (DNS) which we discuss next.

\section{Numerical computations}
\label{sec:numerics}

\begin{figure}
\includegraphics[width=14cm,clip]{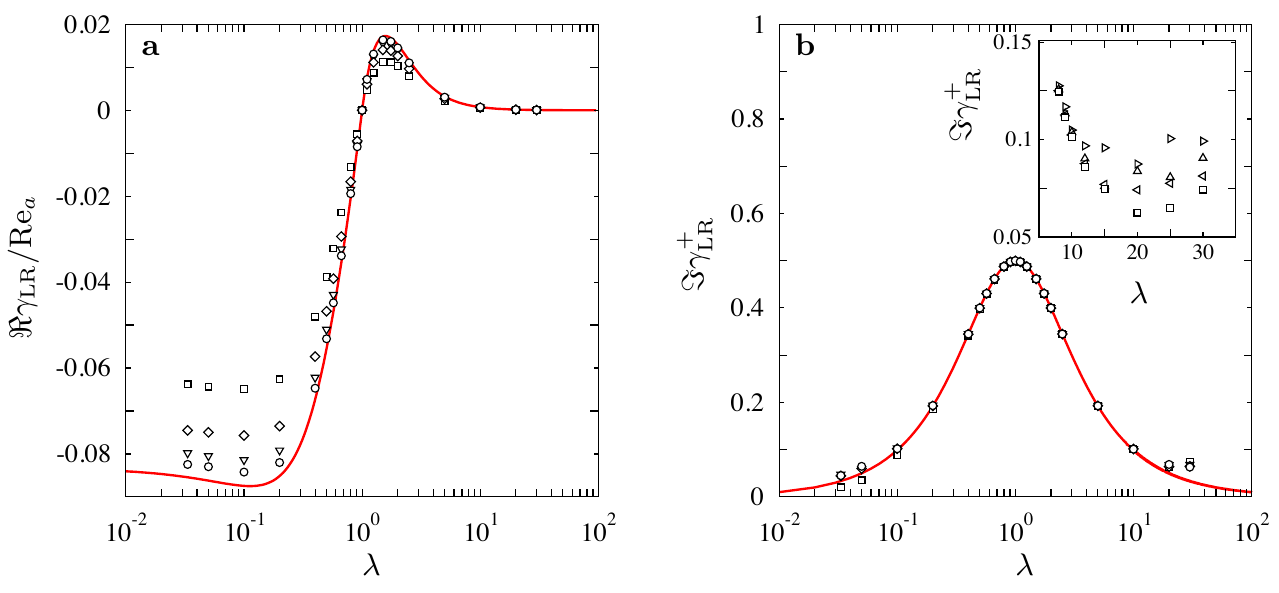}
\caption{\label{fig:gammaLR_rpDep} {(Colour online)}.
{\bf a} Comparison between the analytical result (\ref{eq:gammaLR})
for $\Re\gamma_{\rm LR}$ (solid red line) and numerical results from direct numerical stability analysis (Section \ref{sec:numLR}). Parameters: $\Reys=2.5\cdot 10^{-4}$, $\kappa=0.025$ (circles, $\circ$), $\kappa = 0.05$ (triangles, $\bigtriangledown$), $\kappa = 0.1$ (diamonds, $\diamond$), and $\kappa = 0.2$ (squares, $\Box$). {\bf b} Same comparison for the imaginary part $\Im\gamma_{\rm LR}^+$. The inset shows numerical results for $\Im\gamma_{\rm LR}^+$ for slender prolate spheroids.  Shown are results for $\kappa = 0.2$, $\Reys=2.5\cdot 10^{-4}$, and for different grid sizes in the vicinity of the particle: same resolution as in the main plot (squares, $\Box$), characteristic lengths of the finite elements close to the particle larger by a factor of $1.25$ ($\vartriangleleft$), $1.5$ ($\bigtriangleup$), and $2$ ($\vartriangleright$).}
\end{figure}

We performed different types of DNS of Eqs.~(\ref{eq:particle_eom}) to (\ref{eq:ns}) in a finite
domain with velocity boundary conditions in the shear direction, periodic boundary conditions in the other directions,
and no-slip boundary conditions on the particle surface.
We directly computed the linear stability of the log-rolling orbit using version 4.4 of the commercial software package Comsol Multiphysics\textsuperscript{TM}. As explained below this method could not be used to numerically determine the linear
stability of tumbling in the flow-shear plane. Therefore we used 
lattice-Boltzmann simulations of the particle dynamics to determine the bifurcations of this orbit. To check
the accuracy of the lattice-Boltzmann simulations we also performed steady-state DNS 
using version 9.06 of the commercial software package STAR-CCM+\textsuperscript{TM}.

\subsection{Direct numerical stability analysis of log rolling at finite values of $\Reys$}
\label{sec:numLR}
The eigenvalue solver in version 4.4 of the commercial finite-element software package Comsol Multiphysics\textsuperscript{TM} \cite{comsol_documentation} makes it possible to analyse the stability of the log-rolling orbit as described in this Section \cite{rosen2015c}.
The symmetries of the problem ensure that log rolling exists not only
at infinitesimal but also at finite shear Reynolds numbers.

To determine the linear stability
of this orbit  it is sufficient to account for small deviations of $\ve n$ 
from the log-rolling direction $\ve n = [0,0,1]$, and for the fact that the particle spins around its symmetry axis. Thus we avoid computationally expensive re-meshing around the particle.

The analysis proceeds in two steps. The first step is to find the steady-state solution of Eqs.~(\ref{eq:particle_eom}) to (\ref{eq:ns}) for a given value of $\Reys$, keeping $\ve n$ fixed at $\ve n = [0,0,1]$. 
This determines the angular velocity $\ve \omega$ at which 
the particle spins around its symmetry axis. The second step is to allow for 
infinitesimal deviations of $\ve n$ and $\ve \omega$ from this steady state. 
We use a so-called \lq arbitrary Lagrangian-Eulerian method\rq{} \cite{comsol_documentation} for grid refinement (deformation) close to the particle surface, linearise the resulting dynamics, and determine the eigenvalues of the linearised
problem using the eigenvalue solver in Comsol, which is based on ARPACK FORTRAN 
routines for large eigenvalue problems \cite{comsol_documentation,arpack}. 
The eigenvalue solver provides $N$ eigenvalues $\gamma_1,\ldots,\gamma_N$ closest to the origin in the complex plane, ordered by ascending real parts, $\Re\gamma_{1}>\ldots>\Re\gamma_N$. 

When the shear Reynolds number is small we usually find that $N\!-\!2$ eigenvalues $\gamma_{3},\ldots,\gamma_N$ 
are real (within numerical accuracy) with negative real parts. 
These are fluid modes that decay rapidly as the steady state is approached. In addition there is one 
leading pair of complex conjugate eigenvalues $\gamma_{1,2}$ with largest real part.
This complex pair corresponds to  the linear stability
exponent $\gamma_{\rm LR}^{\pm}$ of the log-rolling orbit. It can have positive
or negative real part, and the imaginary part determines the angular velocity of the particle. 
We must choose $N$  large enough to ensure that this pair is among the $N$ eigenvalues the solver finds. In most cases we find $N=200$ to be sufficient.
At larger values of $\Reys$ it may happen that fluid modes have real parts that are larger
than that of $\gamma_{\rm LR}^{\pm}$, yet they are still real (within numerical accuracy).
When this happens we verify 
that the complex pair describes the stability of the orientational dynamics of the particle
by numerically integrating the dynamics near the steady state.

\begin{figure}
\includegraphics[width=14cm,clip]{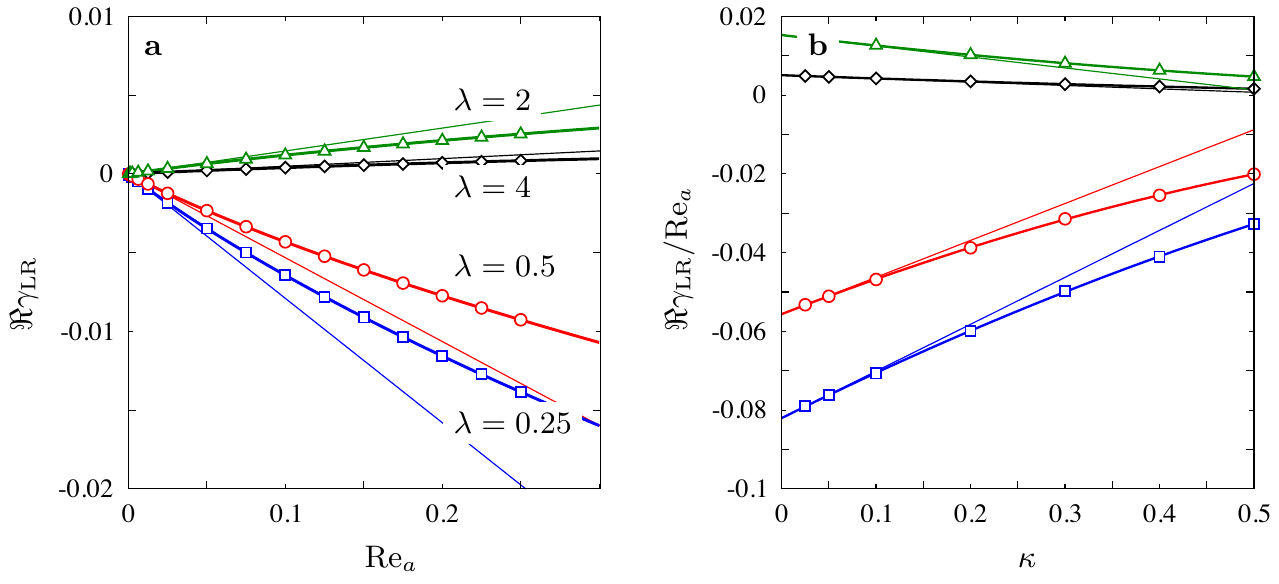}
\caption{\label{fig:gammaLR_Prp2}  {(Colour online)}.
{\bf a} Shows  $\Re\gamma_{\rm LR}$ as a function of $\Reys$ for $\kappa = 0.025$ and for four different values of $\lambda$. The thin solid lines
show the limiting behaviour as $\Reys\to 0$. The thick solid lines show
fits to Eq.~(\ref{eq:fit}). The coefficients are given in Table~\ref{table:c}.
{\bf b} Finite-size corrections to  ${\Re\gamma_{\rm LR}}/{\Reys}$ for $\Reys= 2.5 \cdot 10^{-4}$, and for the same values of $\lambda$ as in panel {\bf a}. The thick solid lines show quadratic fits to the small-$\kappa$ behaviour,
thin solid lines show the corresponding linear $\kappa$-dependence for small values of $\kappa$.}
\end{figure}

In this way we determine $\gamma_{\rm LR}^\pm$ as a function of the particle aspect ratio $\lambda$ 
for different degrees $\kappa$ of confinement, and for different values of $\Reys$. 
Fig.~\ref{fig:gammaLR_rpDep} shows real and imaginary
parts of $\gamma_{\rm LR}^\pm$ as functions of the aspect ratio of the particle, for a small shear Reynolds
number ($\Reys = 2.5\cdot10^{-4}$) and for different system sizes, $\kappa = 0.025,0.05,0.1$, and $0.2$. Fig.~\ref{fig:gammaLR_rpDep}{\bf a} compares the numerical results for the real part of $\gamma_{\rm LR}$ with the theory, Eq. (\ref{eq:gammaLR}).  We observe excellent agreement for the largest system ($\kappa = 0.025$).
This lends support to the analytical results of Refs.~\citenum{einarsson2015a,einarsson2015b,candelier2015}, and also to the numerical linear stability analysis. 
As we reduce the system size ($\kappa = 0.05,0.1$, and $0.2$) we observe increasing deviations from the theory for the unbounded system, as expected. 
{For $\kappa=0.2$}
there
are substantial finite-size corrections.
Fig.~\ref{fig:gammaLR_rpDep}{\bf b} compares numerical results for 
the imaginary part $\Im\gamma_{\rm LR}^+$ with Eq.~(\ref{eq:gammaLR}). Also 
for the imaginary part good agreement between the numerical results and Eq.~(\ref{eq:gammaLR}) is observed for large system sizes, at least for moderate aspect ratios,
$10^{-1} \leq \lambda \leq 10$. As for the real part there are finite-size corrections,
but they are small relative to the $O(\Reys^0)$-term in Eq.~(\ref{eq:gammaLR}).  

Now consider the deviations between the numerical results and theory that can be seen 
in Fig.~\ref{fig:gammaLR_rpDep}{\bf b} for more extreme aspect ratios. In this panel (and also in Fig.~\ref{fig:gammaLR_rpDep}{\bf a}) the size of the finite elements close to the particle surface is  chosen as small as possible given the limited computational memory. But for very large (and also for very small) aspect ratios the resolution is insufficient.  This can be seen in the inset of Fig.~\ref{fig:gammaLR_rpDep}{\bf b}. The inset shows data for $\Im\gamma_{\rm LR}^+$ for $\kappa =0.2$, and for different grid resolutions in the vicinity of the particle.  For moderate aspect ratios the results converge quickly as the mesh size is reduced. But for $\lambda > 10$ we do not obtain convergence, reflecting the limitations of the numerical approach.

Fig.~\ref{fig:gammaLR_Prp2}{\bf a} shows finite-$\Reys$ corrections to $\Re \gamma_{\rm LR}$ for four different values of $\lambda$, for the smallest value of $\kappa$ at which we could reliably compute, $\kappa = 0.025$. Theory \citep{lovalenti1993,candelier2013} suggests that there are $\Reys^{3/2}$-corrections to Eq.~(\ref{eq:gammaLR}) in the unbounded problem ($\kappa \rightarrow 0$). These corrections arise as follows.
The leading-order inertial perturbation of the angular dynamics (linear in $\Reys$) is obtained in terms of the solution of the lowest-order problem, Stokes problem. At finite but small values of $\Reys$ the Stokes solution provides an accurate
description of the fluid velocity in the vicinity of the particle. But at larger distances from the particle (further away than the Ekman length $2a/\Reys^{1/2}$) the actual solutions decay more rapidly than
the Stokes solution. Within the perturbative scheme used in Refs.~\cite{einarsson2015a,einarsson2015b,candelier2015,subramanian2005} this gives rise to a $\Reys^{3/2}$-correction. 
The precise form of higher-order $\Reys$-corrections is not known. We assume that the next order is quadratic in $\Reys$ and compare
the $\Reys$-dependence observed
in the direct numerical simulations with a fit of the form
\begin{equation}
\label{eq:fit}
\Re \gamma_{\rm LR} = a_1(\lambda,\kappa) \,\Reys + a_2(\lambda,\kappa)\,\Reys^{3/2}+a_3(\lambda,\kappa)\, \Reys^2+\ldots\,.
\end{equation}
The values obtained for the coefficients $a_1,a_2$, and $a_3$ are listed
in Table~\ref{table:c}. The data shown in Fig.~\ref{fig:gammaLR_Prp2}{\bf a} and Table~\ref{table:c} are consistent with the existence of $\Reys^{3/2}$-corrections when the system is large enough,
$\kappa \ll \Reys^{1/2}$.
\begin{table}[t]
\caption{\label{table:c} Coefficients $a_1$, $a_2$, and $a_3$
from the fit of Eq.~(\ref{eq:fit}) to
the data in Fig.~\ref{fig:gammaLR_Prp2}{\bf a}, for $\kappa = 0.025$.} 
\mbox{}\\
\begin{tabular}{llll}
\hline\hline
$\lambda$\hspace*{1cm} & $a_1$   & $a_2$ & $a_3$\\\hline
1/4     &-0.0830                                  &\phantom{-}0.0652 &-0.0200\\
1/2     &-0.0566                                 &\phantom{-}0.0482 &-0.0183\\
2       &\phantom{-}0.0155 &-0.0125           &\phantom{-}0.0035\\
4       &\phantom{-}0.0051 &-0.0039           &\phantom{-}0.0008\\
\hline\hline
\end{tabular}
\end{table}
\begin{table}
\caption{\label{tab:c} Coefficients $c_1$, $c_2$, and $c_3$
from the fit of Eq.~(\ref{eq:b}) to
the data in Fig.~\ref{fig:gammaLR_Prp2}{\bf b}. Also given
are the numerical values of $b_4(\lambda)/4$ to which the coefficient $c_1$ should converge as $\kappa \rightarrow 0$ and $\Reys\to 0$. These values are taken from Ref.~\citenum{einarsson2015a} since the
aspect ratios $\lambda = 1/4,1/2,2, 4$ are not small (large) enough
to use the asymptotic formulae given in Table~\ref{tab:1}.}
\mbox{}\\
\begin{tabular}{lllll}
\hline\hline
$\lambda$\hspace*{1cm} & $c_1$ & $b_4(\lambda)/4$ & $c_2$ & $c_3$\\\hline
1/4     &-0.08205          & -0.0820                        &\phantom{-}0.11923&-0.04124\\
1/2     &-0.05564          & -0.0555                        &\phantom{-}0.09373&-0.04521\\
2       &\phantom{-}0.01526& \phantom{-}0.0153\hspace*{10mm}&-0.02784          &\phantom{-}0.01347\\
4       &\phantom{-}0.00510 & \phantom{-}0.0051               &-0.00870          &\phantom{-}0.00367\\
\hline\hline
\end{tabular}
\end{table}

Fig.~\ref{fig:gammaLR_Prp2}{\bf b} shows finite-size corrections to $\Re \gamma_{\rm LR}$ for $\Reys = 2.5\cdot 10^{-4}$ and for four different values of $\lambda$.  Also shown are fits of the form
\begin{equation}
\label{eq:b}
\Re\gamma_{\rm LR}/\Reys= c_1(\lambda) + c_2(\lambda)\kappa + c_3(\lambda)\kappa^2\,.
\end{equation}
The resulting coefficients are given in Table~\ref{tab:c}. We see from Fig.~\ref{fig:gammaLR_Prp2}{\bf b} that 
the fits describe the numerically observed finite-size dependence accurately, but Eq.~(\ref{eq:b}) is just an ansatz. 
Also shown are linear approximations valid at small $\kappa$. We see that the finite-size effects
are to a good approximation linear in $\kappa$ for the data shown for $\kappa \leq 0.1$. Table~\ref{tab:c} shows that the limiting values, {$c_1$,} obtained as $\kappa\to 0$ are in excellent agreement with the theoretical results for the unbounded system.

\subsection{Time-resolved lattice-Boltzmann simulations}
\label{sec:LB}
To analyse the bifurcations of the tumbling in the flow-shear plane we use the lattice-Boltzmann method with external boundary force \cite{Wu2010}. 
To
restrict the computational time, the domain size is set to a maximum of $240$ lattice units.  This allows
us to resolve the particle with at least six fluid grid-nodes along its smallest dimension, with system 
size $\kappa = 0.2$. These choices limit the range of aspect ratios that can be simulated to $\lambda\in[1/8,8]$.  We take $\Reys$ larger than or equal to unity in our simulations. This is because it is computationally very expensive to reach small values of the shear Reynolds number  (as discussed by \citet{rosen2015a,rosen2015b}).

\begin{figure}
\includegraphics[width=8cm,clip]{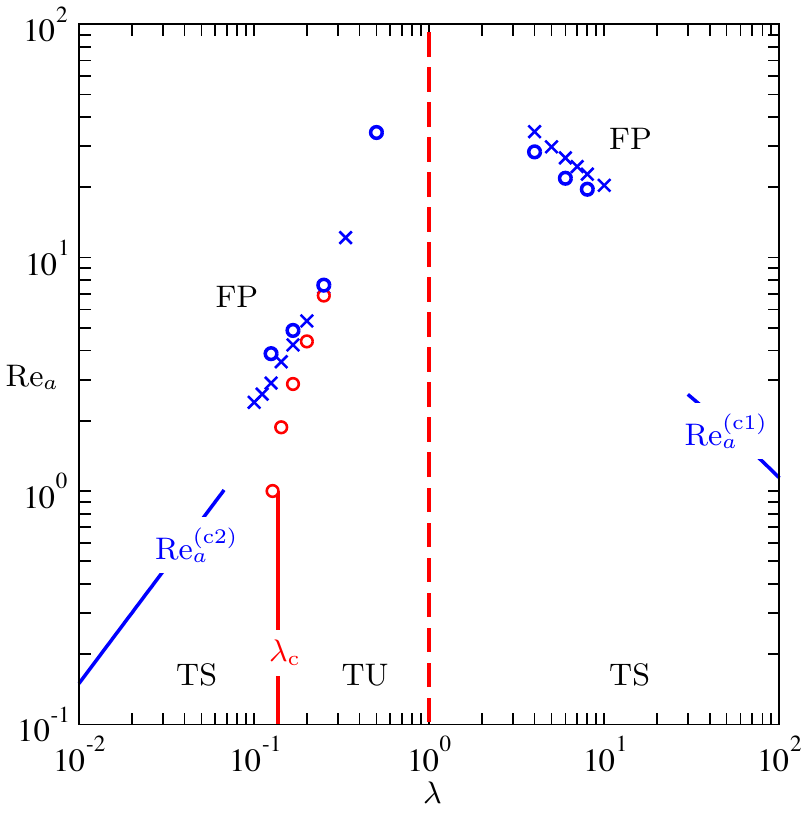}
\caption{\label{fig:lambdac} 
{(Colour online)}.
Bifurcations of the tumbling orbit in the flow-shear plane.
Bifurcation lines derived in Section \ref{sec:theory} for the unbounded system -- Eqs.~(\ref{eq:Rec2}), (\ref{eq:Rec3}), and  (\ref{eq:lambdac}) -- are shown as solid lines. 
The label ${\rm TS}$ indicates that tumbling is stable, ${\rm TU}$ that
it is unstable, and ${\rm FP}$ that the tumbling orbit has bifurcated giving
rise to a fixed orientation in the flow-shear plane.
The dashed line denotes the symmetry line at $\lambda=0$ where the
tumbling orbit changes stability.
Numerical results for the finite system ($\kappa=0.2$) are shown as symbols: circles ($\circ$) denote results from the time-resolved lattice-Boltzmann 
simulations described in Section~\ref{sec:LB},  crosses ($\times$) represent results from the steady-state simulations described in Section~\ref{sec:ss}.  The bifurcations where tumbling in the flow-shear plane changes stability are shown in red, the bifurcations where stable tumbling in the flow-shear plane changes to a stable fixed point are shown in blue. }
\end{figure}

To estimate the critical aspect ratio $\lambda_{\rm c}$ where tumbling changes stability for oblate particles we proceed as follows.  We initialize the particle at rest, close to the tumbling orbit at $\phi=\pi/2$ and $\theta=\pi/2-\delta\theta$ with $\delta\theta = 0.017$. We integrate the dynamics for  aspect ratios $\lambda=1/8, 1/7, 1/6, 1/5, 1/4$, and for $\Reys$ between $1$ and $10$ with unit increments. We determine whether the trajectory tends to tumbling in the flow-shear plane or to the log-rolling orbit, and determine the location of the bifurcation by interpolation.  At $\Reys=1$ we run simulations for $\lambda$ ranging between $0.125$ and $0.160$ with increments of $0.05$ and determine the bifurcation point by linear interpolation.  The results are illustrated in Fig.~\ref{fig:lambdac}. We see that the results  agree fairly well with Eq.~(\ref{eq:lambdac}).  At the smallest value of $\Reys$ simulated with the lattice-Boltzmann code, the transition occurs at $\lambda_c \approx 0.1275$, not too far from the analytical result (\ref{eq:lambdac}) at infinitesimal $\Reys$ for the unbounded system.

Using lattice-Boltzmann simulations we also obtain estimates for $\Reys^{({\rm c1})}$ and $\Reys^{({\rm c2})}$ 
(Section \ref{sec:theory}). This is done by initialising the particle at rest at $\phi=\pi/4$ and $\theta=\pi/2$ for $\lambda>1$ and at $\phi=3\pi/4$ and $\theta=\pi/2$ for $\lambda<1$.  We then determine whether the particle tends to a steady state or continued to tumble, and determine the critical
Reynolds number by linear interpolation.
The results of these simulations are also shown in Fig.~\ref{fig:lambdac}, and are compared with the analytical results for thin disks and rods given by Eqs.~(\ref{eq:Rec2}) and (\ref{eq:Rec3}). We find that the agreement is only qualitative.
This is not surprising since  Eqs.~(\ref{eq:Rec2}) and (\ref{eq:Rec3}) are based on Eq.~(\ref{eq:eom_symmetry}) that is valid only to linear order in $\Reys$ and cannot be expected to 
describe the dynamics at Reynolds numbers of order unity or larger.
We also note that the lattice-Boltzmann simulations were performed for a rather small system, while the analytical results pertain to an unbounded system. Figs.~\ref{fig:gammaLR_rpDep}{\bf a} and \ref{fig:gammaLR_Prp2}{\bf b} show that there are substantial finite-size corrections to the stability exponent of the log-rolling orbit in the finite system, for $\kappa =0.2$. We therefore expect that there are equally important
finite-size corrections to the locations of the bifurcations in Fig.~\ref{fig:lambdac}. But at present we cannot perform lattice-Boltzmann simulations for larger systems with sufficient resolution 
to quantify this statement. 

In order to check the accuracy of the lattice-Boltzmann 
simulations at  $\kappa = 0.2$ we determined the critical Reynolds numbers  $\Reys^{({\rm c1})}$ and $\Reys^{({\rm c2})}$ using an alternative approach, described in the next Section.

\subsection{Steady-state simulations using STAR-CCM+\textsuperscript{TM}}
\label{sec:ss}
We compute the critical Reynolds numbers $\Reys^{({\rm c1})}$  and $\Reys^{({\rm c2})}$  using version 9.06 of the commercial finite-volume software package STAR-CCM+\textsuperscript{TM} \cite{star}.
We choose the same
system size as in the lattice-Boltzmann simulations, $\kappa=0.2$. The particle
orientation is fixed at $\theta=\pi/2$, $\phi\in[0,\pi/2]$ for prolate particles, 
and $\phi\in[\pi/2,\pi]$ for oblate particles. 
For  a given particle aspect ratio $\lambda$ and value of $\Reys$ we compute the steady-state torque on the particle. If the torque vanishes, the chosen particle orientation is a fixed point for the given parameters.
A fixed particle orientation makes it possible to use a very fine local grid around the particle. 
For different choices of  $\phi$ we find  critical Reynolds numbers where the steady-state torque vanishes. 
The minimum of this critical Reynolds as a function
of $\phi$ gives $\Reys^{({\rm c1})}$ or $\Reys^{({\rm c2})}$,
for prolate and oblate particles respectively. 
The corresponding results for $\Reys^{({\rm c1})}$ and $\Reys^{({\rm c2})}$ are also shown
in Fig.~\ref{fig:lambdac}. We conclude that the lattice-Boltzmann simulations slightly underestimate the 
critical value $\Reys^{({\rm c1})}$, while they slightly overestimate $\Reys^{({\rm c2})}$.

\section{Conclusions}
\label{sec:conc}
Using numerical linear stability analysis we computed the stability of the log-rolling orbit of a neutrally buoyant spheroid in a simple shear at small $\Reys$. For infinitesimally small $\Reys$ in the unbounded system this problem was recently solved for arbitrary aspect ratios using perturbation theory in the shear Reynolds number. The fact that both calculations agree in the limits $\Reys\to 0$ and $\kappa\to 0$ (unbounded system) lends support to the analytical calculations \cite{einarsson2015a,einarsson2015b,candelier2015}, but also to the numerical linear stability analysis described in the present article.  In the limit of large system size ($\kappa\to 0$) we found that there are corrections 
to the analytical result for the exponent $\Re\gamma_{\rm LR}$ that are consistent with terms of order $\Reys^{3/2}$.  We also investigated finite-size corrections to $\Re\gamma_{\rm LR}$ at small $\Reys$, and found that they are substantial. It would be of interest to calculate both finite-$\Reys$ and finite-size corrections to $\Re\gamma_{\rm LR}$
by extending the method used in Refs.~\citenum{einarsson2015a,einarsson2015b,candelier2015}. 

We did not  investigate the stability of the tumbling orbit with numerical linear stability analysis because the required re-meshing is computationally very expensive. Instead we 
studied the stability of tumbling in the flow-shear plane using lattice-Boltzmann simulations.
We tracked the bifurcation line between stable/unstable tumbling for thin oblate spheroids (solid red line in Fig.~\ref{fig:lambdac}) down to as small values of $\Reys$ as we could reliably achieve and found that the transition occurs at $\lambda_{\rm c}\approx 0.1275$ at $\Reys=1$, in 
fair agreement with the theoretical prediction $0.137$.

Finally we determined for which values of $\lambda$ and $\Reys$  tumbling in the flow-shear plane bifurcates
to a fixed point, using lattice-Boltzmann simulations, and also by numerically computing steady-state
torques using STAR-CCM+\textsuperscript{TM}. The two numerical procedures give results that
are in fairly good agreement with each other, yet the agreement with the analytical results
(\ref{eq:Rec2}) and (\ref{eq:Rec3}) is only qualitative. 

Detailed analysis of the lattice-Boltzmann dynamics near
the bifurcation at $\Reys^{({\rm c2})}$ reveals the phase-space topology near the bifurcation at moderate
Reynolds numbers ($\Reys^{({\rm c2})}\approx 7.8$ at $\lambda = 1/4$), see Fig.~3(d),(e) in Ref.~\citenum{rosen2015b}.
For $\lambda = 1/4$  a second transition occurs at $\Reys^{({\rm c3})}\approx 5$ 
where the log-rolling orbit changes from stable spiral to stable node. 
Eq.~(\ref{eq:eom_symmetry}) also exhibits this transition. But since Eq.~(\ref{eq:eom_symmetry}) is valid to linear order in $\Reys$ the bifurcation can only be analysed in the limit $\lambda\to 0$. We find that the two transitions occur in reverse order: the tumbling$\to$fixed point bifurcation occurs before the spiral$\to$node transition  as the shear Reynolds number is increased. 
There are several possible explanations for these subtle differences.
They could be due to higher-order $\Reys$-corrections to Eq.~(\ref{eq:eom_symmetry}) such
as the $\Reys^{3/2}$-corrections alluded to above. 
But we have also observed (not shown) in the numerical simulations of the bounded system that $\Reys^{({\rm c3})}$ increases as $\kappa$ becomes smaller. In the limit of $\kappa\to0$ we except 
that the order of the transitions  agrees with the prediction
for the unbounded system.
In summary we can conclude that the results of our numerical computations agree well with the theoretical
predictions at infinitesimal Reynolds numbers: we find
excellent agreement for the stability exponent of the log-rolling orbit,
and the bifurcation of the tumbling orbit for thin oblate particles 
occurs in both theory and simulations, at similar values of $\lambda_{\rm c}$. But there are a number of subtle
differences between theory and simulations at larger Reynolds numbers. At present we cannot
reliably perform lattice-Boltzmann simulations at much smaller Reynolds numbers than those 
shown in Fig.~\ref{fig:lambdac}, and it is very difficult to perform such simulations at still smaller values of $\kappa$.  Therefore it would be of great interest to extend the analytical calculations to include $\Reys^{3/2}$-corrections, and to account for finite-size effects.

\acknowledgments{JE and BM acknowledge  support by Vetenskapsr\aa{}det and by the grant \lq Bottlenecks for particle growth in turbulent aerosols\rq{} from the Knut and Alice Wallenberg Foundation, Dnr. KAW 2014.0048. 
TR and FL acknowledge financial support from the Wallenberg Wood Science Center (WWSC), Bengt Ingestr\"{o}m's Foundation, and \AA Forsk (\AA ngpannef\"{o}reningen's Foundation for Research and Development).
{Support from the MPNS COST Action MP1305 \lq Flowing matter\rq{}
is gratefully acknowledged.}
The simulations were performed using resources provided by
the Swedish National Infrastructure for Computing
(SNIC) at the High Performance Computing Center
North (HPC2N) and at the National Supercomputer Center
(NSC) in Sweden. We also acknowledge the help from E. Sundstr\"om
in setting up the simulations using STAR-CCM+.}

\end{document}